# The Morality Game: An online multiplayer platform to standardize, expedite, and expand research on cooperation

Gregory N. Stanley, Alan Yang, Liam Tsimhoni, Roujia Wang, Ekdatha Arramreddy, Rohan Desai, Shulin Pan, Michael Ahn, Vijairam G. Moorthy, William Mathias, Mai Xu, James Kessler, Yahan Xing, Yang Li, Duoming Bian

***Abstract***

This paper presents the Morality Game, a platform designed to standardize and accelerate research on cooperation and morality through game theory-based experiments. The Morality Game functions as a video game for science, a hub for economic game research, an open-access data repository, and a tool for expediting the research process. It allows researchers to launch customized online multiplayer experiments with zero coding, using game trees to simulate moral dilemmas. The platform automates participant payments, data collection, and analysis, promoting replication and transparency.

This paper details the platform's architecture, emphasizing its capabilities for standardizing research methods, unifying data, and enabling rapid aggregation and comparison of results. The Morality Game leverages dynamic, self-correcting game trees to generate well-controlled, abstract experiments that can represent any social scenario. Participants interact through a responsive user interface, making the experiments intuitive and engaging.

Researchers can configure experiments through a user-friendly dashboard, specifying various parameters and utilizing pre-created or auto-generated game trees. The platform supports nested belief representation and incorporates artificial agents with customizable traits. Plans include integrating social networking features, enhancing emotional expression capabilities, and expanding the platform's reach to remote small-scale societies to test the ecological validity of findings.

By evolving into an integrated ecosystem that supports the entire research lifecycle, the Morality Game aims to foster collaboration, enhance data accessibility, and ultimately increase cooperation. This paper outlines the platform's current features, architectural details, and future directions, demonstrating its potential to advance cooperation research.

***Highlights***

- The Morality Game is an online platform for behavioral cooperation research
- Researchers launch multiplayer economic-game experiments with zero coding
- Configurable game trees model any social dilemma with human or artificial players
- Automated payments, data, and analysis support transparent replication
- Every study is a condition in one cumulative, comparable design space

***Corresponding author***

Gregory Stanley - https://orcid.org/0000-0003-1520-9010. E-mail address: gregstan@umich.edu. Institution: University of Michigan, Department of Psychology, Cognition and Cognitive Neuroscience, 530 Church Street, Ann Arbor, MI 48109-1043, USA.

***Author contributions***

G.N.S. conceived and designed the Morality Game, developed the platform, wrote the large majority of the codebase, wrote the manuscript, and led the following team of research assistants. A.Y. wrote the first version of the game engine, built many of the platform's graphical components, and deployed the platform to production. L.T. contributed to the platform's theoretical development and to the matching function. R.W. contributed to web application development. E.A. contributed to data infrastructure. R.D. established the system for crediting participants via SONA and paying participants via PayPal. S.P. contributed to the platform's foundational architecture during early development. M.A. contributed to early foundational development, including the matching function. V.G.M. developed the backend WebSocket interface through which users interact in real time. W.M. contributed to the platform's interaction and visual systems. M.X. contributed frontend and backend components of the game engine. J.K. worked on theory and the bots. Y.X. helped develop the game interface. Y.L. developed the experiment dashboard in React. D.B. developed the single-player online experiment that was the direct predecessor to the Morality Game, establishing the web-based foundation from which the platform grew.

# 1. Overview

### 1.1 Platform Goals

The Morality Game is a video game for science, a hub for economic game research, an open-access data archive, and a tool for standardizing and propelling research on cooperation and morality. It is a platform where researchers launch customized online multiplayer experiments and subjects participate for credit or payment based on their decisions. Experiments consist of rounds where human and artificial players match and reveal their values in moral dilemmas represented by game trees from game theory. Rather than coding entire experiments, researchers login to the site, configure the conditions of their study, and click 'deploy'. The system automatically executes the experiment, credits or pays participants via PayPal, and returns the data, formatted, analyzed, and stored in a centralized database so that others may benefit from this knowledge. Researchers can simulate any experiment with programmable artificial agents varying in traits like selfishness, risk aversion, and reasoning depth.

The Morality Game is a research tool for replication, unification, and speed. The replication crisis in psychology is exacerbated by the difficulty in reproducing experiments and the lack of transparency (Ioannidis, 2005; Diener & Biswas-Diener, 2016). Traditional methods sections often lack the granularity needed for precise replication, especially for novel methods. The Morality Game addresses this by allowing researchers to replicate experiments through a 'replicate' button, which imports all settings of an existing study into their dashboard, enabling immediate deployment. The settings are stored as JSON files, capturing the entire methodological blueprint for exact replication. Additionally, the Morality Game enhances transparency by making all methods and results publicly accessible[1].

In addition to aiding replication, by standardizing research methods and unifying the data, the platform accelerates the creation and dissemination of knowledge. Increasing the rate of scientific progress is a universally acknowledged goal. However, the current state of psychological research is fragmented, with data dispersed across multiple studies, each employing different procedures. This fragmentation poses a significant challenge for conducting meta-analyses, as researchers must reconcile disparate techniques and results to draw meaningful conclusions.

The Morality Game addresses this issue by centralizing data into a uniform format, allowing for rapid aggregation and comparison of results. Each study's outcomes are assimilated into a comprehensive database. This approach eliminates the tedium and potential errors associated with data preprocessing, as all data types, column names, and handling of missing values are standardized across the platform.

Methodological unification aids theoretical unification. With all data stored in one place, theoreticians and modelers are better able to identify overarching patterns[2].

---

[1] Planned Feature: Researchers will be able to embargo data for a limited time.

[2] Planned Feature: Furthermore, the Morality Game platform will foster a collaborative environment where researchers can communicate, share ideas, and build upon each other's work (social networking features not yet added, except for data sharing).

### 1.2 Game Trees

Game trees are the fundamental unit of the platform, mathematical structures from Game Theory that can represent the abstract dynamics of any social or moral dilemma (Von Neumann & Morgenstern, 1947; Osborne & Rubinstein, 1994). Like multi-agent decision trees, these structures are composed of players, payoffs, probabilities, states, actions that link states, and players' beliefs about these elements. Nodes represent states of the dilemma and lines connecting nodes represent one-way transitions between states. When a player or players can choose, these edges represent their options. When players cannot choose, outcomes are determined by chance via probabilities associated with these state changes. Some states contain payoffs, which are rewards like points for each player, and which translate into payments in paid studies. Games begin at the root node, at the top of the screen, and progress towards leaf nodes, at the bottom of the screen. Games only ever progress downwards—state transitions are always one-time and unidirectional because game trees do not have loops. Thus, game trees represent the progression of time, downwards on the screen, and branching possibilities: what was, what is, what might be, and what would have been, could have been, should have been, etc.

The abstract structure of game trees facilitates experimental control, simplifies computational modeling, and allows representation of diverse social scenarios. By condensing complex real-world situations into a limited set of abstract components that can be systematically linked and combined, game trees offer broad applicability. Consequently, the Morality Game platform can accommodate a wide range of experiments within behavioral game theory and related fields.

However, ecological validity and user-understanding need not be traded for experimental control, mathematical precision, and abstract universality. The real-world features that have been subtracted from the on-screen game trees can be added back. The Future Directions section details my intent to incorporate features such as the ability to communicate with other players, express emotions on player avatars, and maybe even with webcams someday. Indeed, economic game experiments run purely on the platform are intended to be completed in parallel with field studies to test the generalizability of results. Moreover, participants readily understand the dilemmas depicted by trees, mapping social emotions like trust and the feeling of being betrayed to the trees. This is all explained in the tutorial and made intuitive by a responsive user-interface.

### 1.3 User-Experience: Participants

Participants sign up for the platform by entering a username, password, and email address, creating a permanent profile. Each participant is assigned an avatar with a simple geometric shape and a randomly generated color scheme. From the main lobby page, they can click 'Enter Tutorial' to read a general consent form that applies to all experiments and then learn how to participate.

The tutorial mimics the user interface of the official experiments, such as the one depicted in Figure 1, and provides practice opportunities. An overlay with text and YouTube videos explains the task. Participants see their avatar and the avatar of a pre-programmed counterpart located on a node of the on-screen game tree. The current node glows to indicate the present moment. If the participant is the current chooser, they see their options as lower nodes with arrows leading towards them, one of which is faintly colored to match their avatar. This colored arrow points towards the option they will choose if they finalize their selection by pressing the spacebar. Once they press the spacebar, the arrow pointing to their choice becomes more vibrant, giving clear feedback.

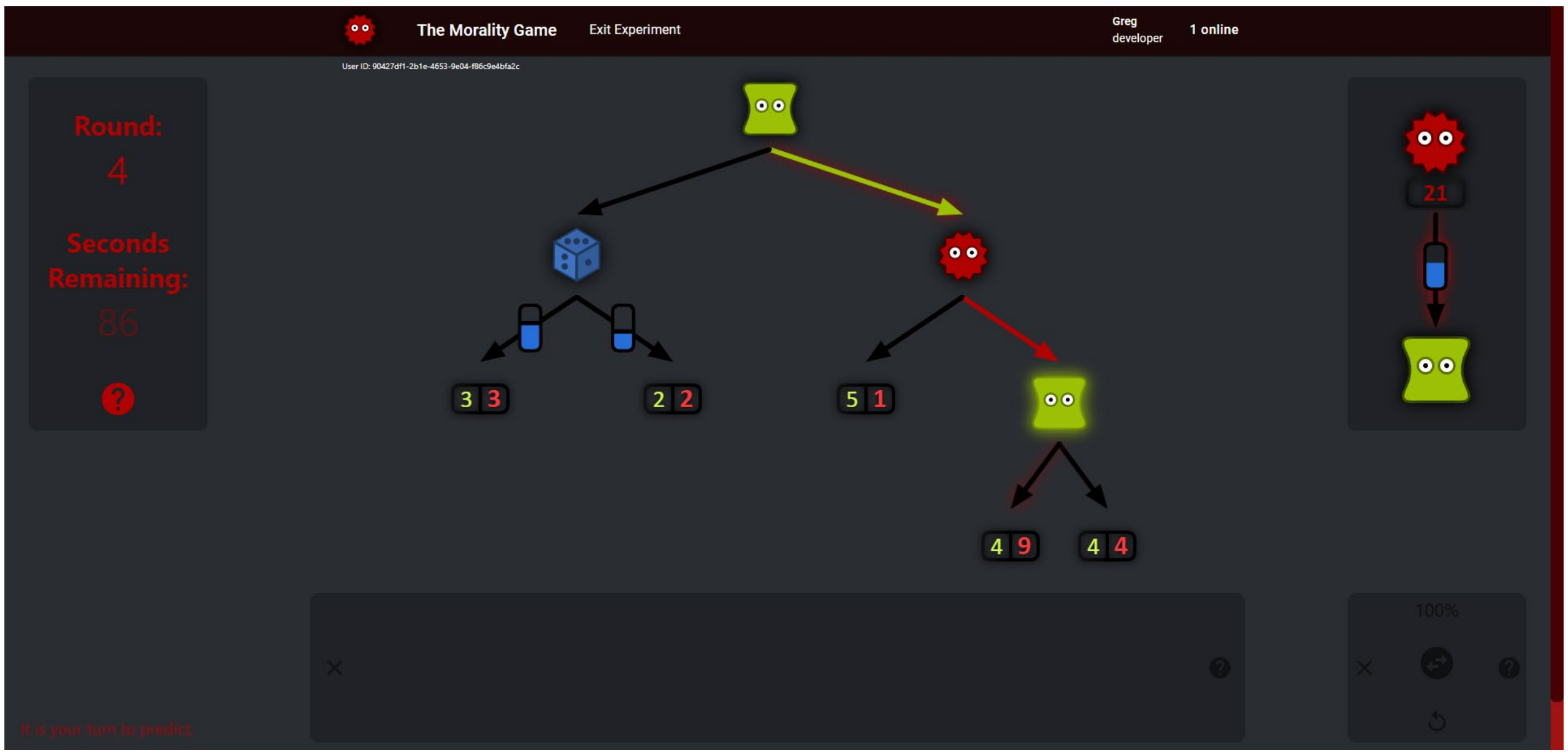

Figure 1: Experiment User-Interface:

The above screenshot is taken from a pilot study. My avatar is the red sun, and my counterpart is the yellow hourglass. The right-side panel indicates my cumulative payoff total of $21$ and a $0.6$ probability of being matched with the yellow counterpart in the next round, as shown by the blue mercury level being $60$ - $70\%$ up the left thermometer.

It is round $4$ and I, Greg, have $86$ seconds remaining to predict whether my counterpart will choose the payoffs of $(4, 9)$ on the left or $(4, 4)$ on the right. The current node is identifiable by its glowing effect. The faint red shadow surrounding the left arrow pointing to the payoffs $(4, 9)$ indicates my current prediction, which will be finalized if I press the spacebar. The topmost arrow pointing to the right has a yellow interior, indicating my counterpart's previous choice, and a red exterior, indicating my previous prediction, which was correct. Similarly, the red arrow on the right side under my avatar shows that I previously chose the right-side option. Predictions are private and are not visible to counterparts, which is why there are no yellow shadows around my previous choices.

The die icon on the left side of the tree represents a chance node, with approximately a $2/3$ probability of resulting in the $(3, 3)$ outcome and approximately a $1/3$ probability of resulting in the $(2, 2)$ outcome. However, the probability mercury is partially covered by a thin black cover, indicating that these probabilities are approximate, also known as ambiguous. This ambiguity cover can range from nonexistent to fully covering the thermometer.

The grey rectangle under the tree is the response pad, where participants can respond by shifting and clicking with a mouse or by swiping and tapping with their fingers if they prefer using touch input over the keyboard. The box in the bottom-right corner allows users to resize the tree as needed.

Before making a choice, participants can scroll through their options: pressing "A" to scroll left and "D" to scroll right, changing the color of the arrow. These changing arrow colors provide 'pre-choice' and 'post-choice' feedback. Once all players have submitted their selections, the game progresses to a new node, which then glows to indicate the state transition. While payoffs can be associated with intermediate nodes, games often progress towards payoffs at a terminal node, called a leaf node. The payoffs are displayed in an array of colored numbers, with the colors matching each player's avatar, and the user's payoffs are visibly larger and bold.

A countdown timer displays the remaining seconds to select an option before the response is considered 'abdicated', resulting in a deduction from the player's cumulative payoff total, which is also displayed on-screen. Participants are instructed that they will be ejected from the study after two minutes of inactivity or if they switch browser tabs. After two minutes of inactivity, a modal with a 30-second countdown timer appears warning participants of their imminent ejection, which they can prevent by moving their cursor or pressing any button.

For those familiar with game theory, the platform supports the display of information sets and simultaneous choices[3]. Chance nodes are states that transition into future states based on probabilities, represented by a die (dice) icon. The probabilities are shown as levels of mercury in thermometers on the arrows pointing to the child nodes, with higher mercury indicating a greater probability of transitioning to that node. Although an advanced concept introduced in tutorial 2, the exact height of the probability mercury can be occluded by a black cover, making the probabilities ambiguous. This degree of ambiguity is expressed by the percentage of the thermometer height covered. While this might seem complex, studies have shown that even chimpanzees can modulate their responses to probabilities made ambiguous in this way (Hayden, 2010; Rosati, 2011).

When the counterpart is the chooser, participants predict this choice rather than simply waiting. Predictions are entered the same way as choices, but the feedback is shown as an avatar-colored glow or shadow on the exterior of the arrows, contrasting with the interior arrow-color in pre-choice and post-choice feedback. Predictions are private and do not affect the games but provide valuable data and keep the game interesting by encouraging participants to consider the minds of their counterparts.

Once all players submit their responses or abdicate by running down the clock, the round ends and the next round begins, unless the experiment ends. Each round, participants are matched with the same or new counterparts—bots or real humans—according to matching probabilities set by the researcher. Participants see the probability of being matched with their current counterparts in the next round. Some participants may choose to be more cooperative with those they expect to see again, while others may not consider this factor.

[3] Planned Feature: Tutorials will be tiered where the system unlocks experiments once participants complete various tutorials.

In each of these games, participants encounter choices involving varying degrees of cooperation, fairness, and trust. However, these scenarios are more accurately described as moral dilemmas rather than conventional games. Although the term "Morality Game" and its interactive audiovisual design might suggest a recreational context, the instructions explicitly present the tasks as moral decisions with potential implications for real human counterparts who value the payoffs involved. As in real-life situations, participants face decisions without predefined objectives, guided primarily by their individual moral judgments.

**1.4 User-Experience: Researchers**

Researchers sign up by entering a special password, “utility.” Once signed up, they can perform all participant functions and have additional capabilities to configure studies and access results. The lobby page features a table of experiments, each row displaying a unique experiment code, the responsible researcher’s ID, and various action buttons depending on the user’s role (guest, participant, researcher, or developer) and the experiment's state (inactive, open to join, ongoing, or finished). Researchers see additional buttons like ‘results’ to view past experiment results and ‘replicate’ to import another study’s settings into their dashboard.

Upon signing up, each researcher receives a dedicated row in this table with a unique code and default experiment settings. Clicking ‘configure’ takes the researcher to their dashboard, shown in Figure 2, where they can customize their experiment's settings, without writing a line of computer code.

Experimenters configure their study on this page. For instance, the top left box includes settings for the number of rounds and number seconds given to respond at each node and the top middle box includes a dropdown menu of game trees and options to determine the proportions of $n$-player games ($n = 1, 2, 3…$). All the settings on the bottom row represent probability distributions. Clicking on these settings opens an overlay where the researcher can establish the type and parameters of the distribution. The left sidebar includes buttons to save settings, reset to default settings, download a JSON file of the settings, simulate the study, pilot the study, or deploy it officially.

These settings are saved in a downloadable JSON file, which seeds the experiment, instructing the system on how to execute the study. Researchers can adjust several crucial settings, such as the number of humans versus artificial players, whether participants can join mid-experiment, and whether the experiment starts at a specific time or once a desired number of humans join. They can predetermine the exact number of rounds or set an $s$-shaped probability distribution over the number of rounds, with the experiment proceeding or terminating based on a random number compared to the continuation probability at each round.

Researchers can create different levels of time pressure by setting the response time for each node. By default, one second is added for each descendant node, but this can be adjusted. They can also set rest breaks between levels of the tree and establish an abdication penalty for failing to respond in time. The matching function allows researchers to set a probability distribution over player matches, ensuring these probabilities are maintained. Researchers can also set the distribution of group sizes (e.g., $\%$ of 1-player, 2-player, and $n$-player games).

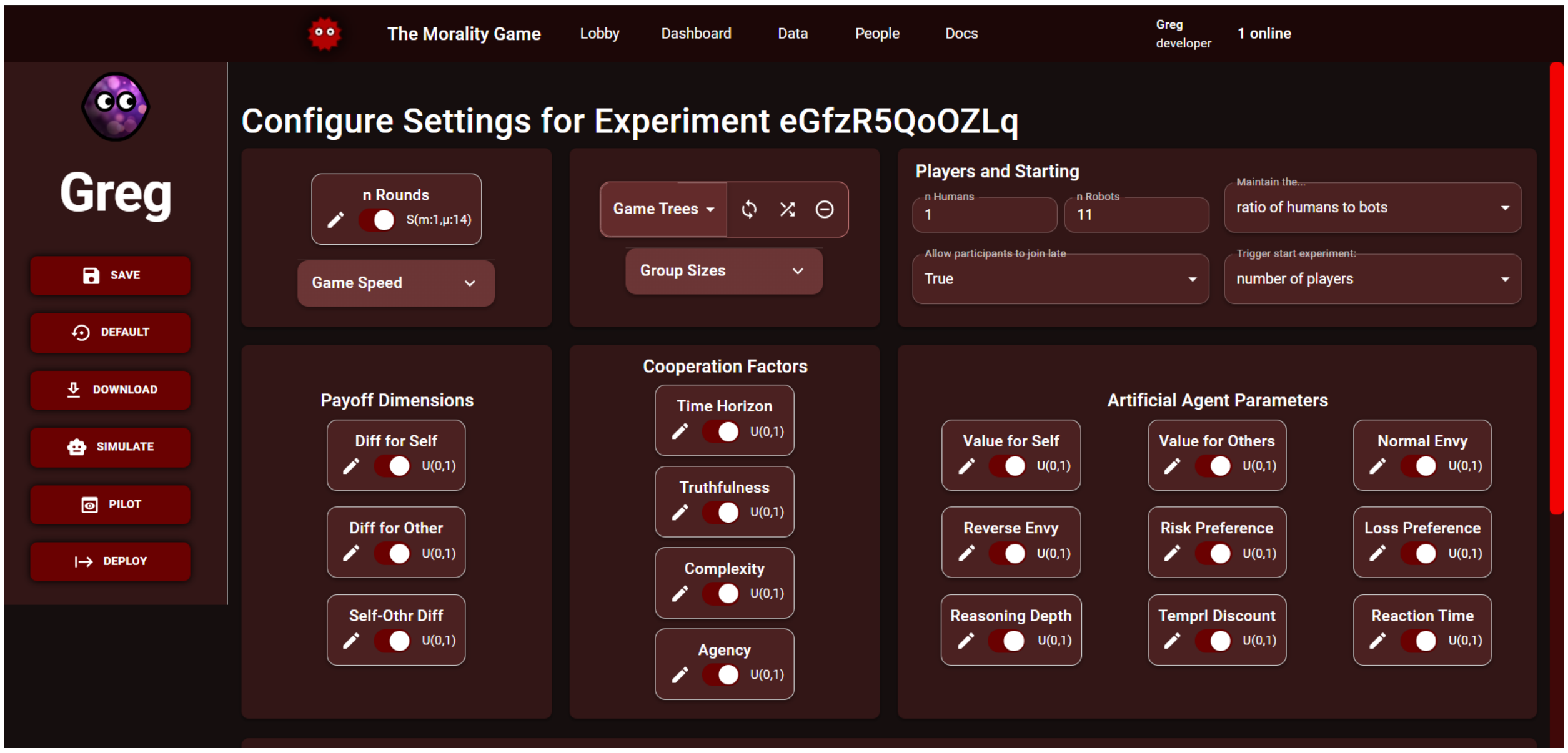

Figure 2: Researcher Dashboard:

This is the user-interface where researchers configure experiments. Starting with the top panel from left to right, settings for the experiments include the number of rounds, the number of seconds participants have to respond at each choice point, the game trees, the number of players per game, the number of humans versus bots, and various starting conditions, such as if the experiment begins at a specific time or once the desired number of human players is reached. The bottom panels for payoff dimensions and cooperation factors determine the game trees if the researcher opts not to select from a list of trees that have already been created. The bottom right panel includes settings that establish the 'personalities' of the bots, including their social preferences, reasoning depths, and reaction times. The leftmost panel features buttons for saving the settings, reverting to default settings, downloading the settings, simulating the experiment with all bots, piloting the study, or deploying the study officially. Although not visible in the image, a lower panel includes widgets for uploading consent/debriefing forms, completion URLs, and setting up payments via PayPal.

Researchers can choose from pre-created game trees or generate game trees automatically based on probability distributions over mathematical properties relevant to cooperation, such as agency, entropy, payoff disparities, and truthfulness. Truthfulness measures the similarity between expected payoffs in the true dilemma and those presented on-screen. Detailed explanations of these processes are provided in the sections on Tree Generation and Cooperation Factors and Beliefs and False Beliefs.

As described in section 4.4 about Artificial Agents, if the study involves bots, researchers can configure the bots' traits, such as value for self, value for others, normal envy, reverse envy, risk preference, loss preference, reasoning depth, temporal discounting, and reaction times. These traits feed into the bots' utility functions, determining their choice probabilities. Researchers can specify these settings probabilistically, with an overlay allowing them to set distribution types (uniform, linear, normal, sigmoid, or point) and parameters (mean, standard deviation, etc.). Probabilistic settings make experiments robust to varying participant numbers, but fixed conditions can also be predetermined.

Researchers can upload consent forms and custom instructions. They can also provide inputs for participants to click a SONA completion URL for automatic crediting. Researchers can set the monetary value of payoffs, determining participant compensation via PayPal.

The researcher dashboard sidebar contains buttons: save, default, download, simulate, pilot, and deploy. The ‘save’ button saves settings, while ‘default’ resets the dashboard to default settings. The ‘download’ button downloads the configuration JSON file. The ‘simulate’ button runs the study with bots, useful for visualizing results and creating data pre-analysis plans. The ‘pilot’ button launches informal studies to troubleshoot issues before formal deployment. Deploying an official study locks the settings, saves the data permanently, and generates a new dedicated row in the experiments table.

Figure 3 describes how researchers can view the progress of ongoing studies.

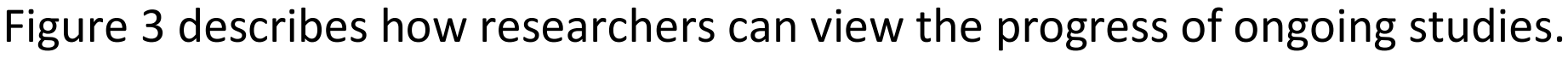

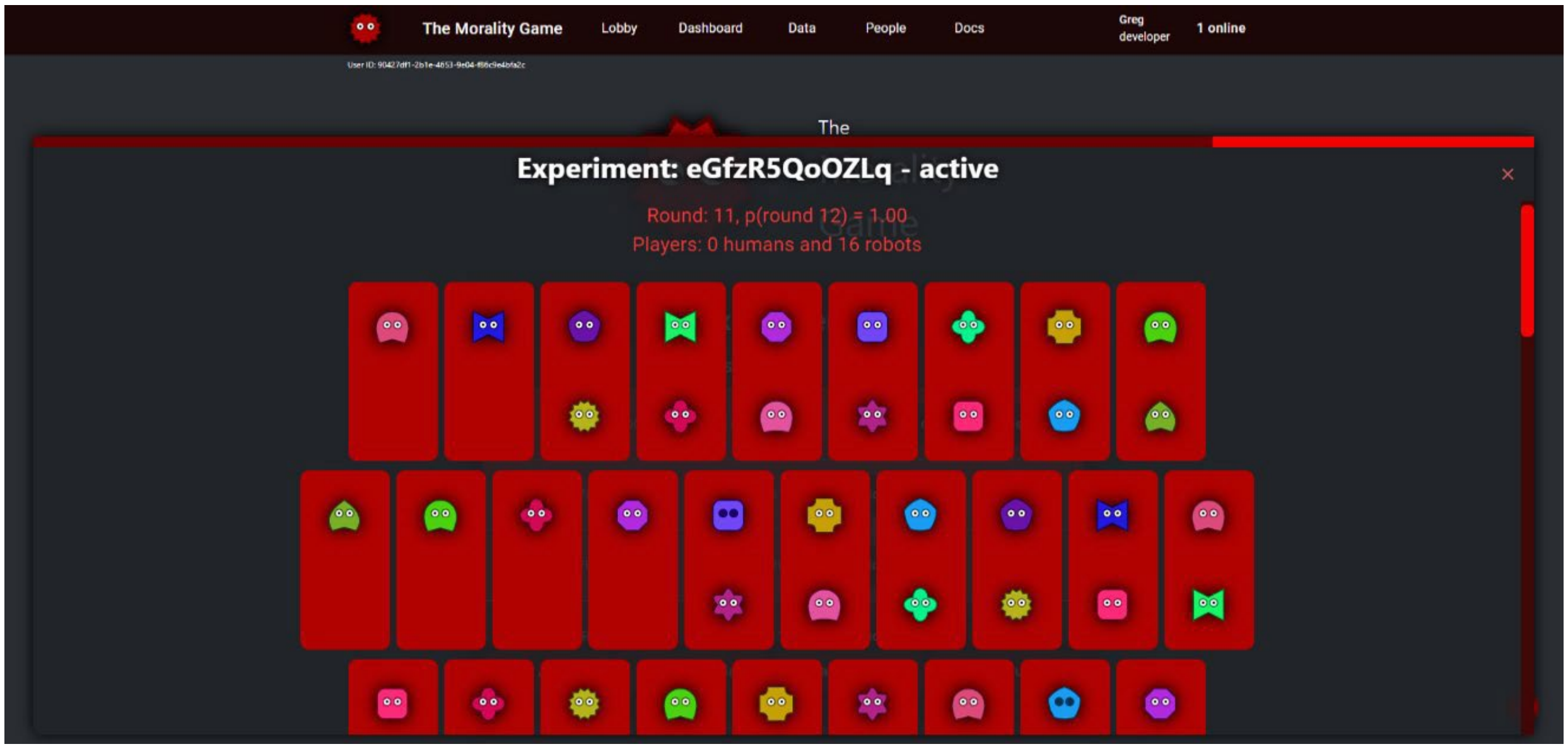

Figure 3: Experiment Progress.

The figure above shows the user interface for a researcher inspecting the progress of their active study. This simulation involves 16 bots, but the interface is identical for human participants. Each row represents a round, each cell represents a game room, and each avatar represents a player. All rounds start and stop synchronously.

The most recent round, shown at the top, has two one-player rooms and seven two-player game rooms. The purple pentagon and yellow sun avatars, as well as the light purple rounded square and red-pink star avatars, were matched in back-to-back rounds, while the other players were paired with new counterparts or played alone in the previous round.

Researchers can hover their cursors over the rooms and avatars to view additional details as hover text. Future versions of the Morality Game will allow researchers to inspect the specific choices made on each game tree and analyze various metrics as they evolve across rounds.

From the experiments table, researchers can access past study results via the ‘results’ button, which leads to the data gallery page. This page presents interactive data visualizations, including line graphs of payoff totals per player and a 3D phase space diagram modeling participant progress through payoff space each round.

## 2. Cooperation Space

The Morality Game exists to increase cooperation. Promoting transparency, reproducibility, collaboration, unification, and efficiency are means to this end. As detailed in section 3, "The Space of Experiments," Morality Game experiments and their constituent game trees vary along dimensions relevant to cooperation, termed cooperation factors. These factors include the degree to which payoffs are aligned, the extent of player control over outcomes, the accuracy of players' beliefs about the game, the transparency of choices, the complexity of scenarios, and the predictability of outcomes.

Multiple dependent variables are measured, such as prediction accuracy, reaction times, and payoff totals, but the cooperation rate is the most interesting dependent variable. The cooperation rate is the mean cooperativeness of all choices at a given point, within a game tree, or across an experiment, depending on the desired resolution. Cooperativeness is essentially the difference between the sum of expected payoffs for all players in the chosen option and the maximum possible expected payoff sum in the set of options the chooser is aware of. A perfect cooperativeness score is zero, indicating that the chooser selected the optimal option for all players.

Each choice can be mapped onto coordinates along the cooperation factors, positioning it within a multidimensional cooperation space. By averaging the cooperativeness across all data points in this space, we create a terrain with peaks where cooperation flourishes and valleys where it fails. This endeavor is akin to navigating a "moral landscape," striving to scale ever higher peaks of cooperation (Harris, 2010).

However, this utilitarian journey to maximize cooperation is only meaningful if Morality Game results predict cooperation rates in real-world contexts. Therefore, it is crucial for researchers to develop parallel lab and field studies to test the generalizability of Morality Game data. If ecological validity is found to be lacking, the challenge lies in incorporating additional factors into experiments that the current set of conditions does not yet capture. Conversely, when Morality Game results accurately predict real-world cooperation rates, these findings can inform public policy, institutional design, and interpersonal relationships. The Morality Game will have achieved its goal when it produces actionable knowledge for promoting cooperation.

The remainder of this paper includes the following sections:

- **3. The Space of Experiments**: Explains how all studies are conditions in an ongoing mega-study.
- **4. Platform Architecture**: Describes the inner workings of the platform.
- **5. Future Directions**: Outlines plans for expanding the Morality Game.

**Advantages of the Morality Game**

**1. Participant Recruitment**

• Automated participant crediting via SONA and payment via PayPal.
• Eventually, the platform will eliminate the need for external recruitment platforms.
• Potential to build a large pool of return participants due to the engaging and paid studies.
• Access to diverse participant populations outside universities, reducing WEIRD sample bias.
• Benefits from network effects where the platform's value and data richness scale exponentially with additional participants.

**2. Bots and Simulations**

• Supports human-human, human-bot, and bot-bot experiments, making it suitable for experimentalists and modelers alike.
• Identical data and procedures for conducting experiments and running simulations.
• Ability to simulate data before conducting experiments to create a rigorous analysis plan or for the intrinsic value of the simulations.
• Bots can enact desired experimental conditions, such as always defecting or reasoning at a specific depth.
• In pilots, modelers can interact with the bots they programmed, offering a hands-on experience of their models.

**3. Flexibility and Accessibility**

• Researchers can test experiments informally at no cost with pilot experiments.
• No programming knowledge required; studies can be launched entirely from a GUI.
• Stimuli automatically adapts to screen sizes, and the interface is sleek and appealing.
• Works on any device through a browser, with keyboard, mouse, or touchscreen input.
• Built-in tutorial feature that instructs participants, with options for custom instructions.
• Frequently-asked questions panel available mid-experiment for participants to reference.
• Conduct lab, field, or online studies on desktops, tablets, and smartphones (coming soon).

**4. Data Management and Visualization**

• Automatic visualization of results in researcher's data gallery.
• Standardized and open-access data for ease of aggregation and analysis.

**5. Experiment Monitoring and Customization**

• Matching of participants according to probabilities specified by the researcher.
• Real-time inspection of ongoing experiments, allowing for immediate feedback.
• Penalizes and ejects inactive participants to prevent them from stalling the study.
• Minimal consent form uploads due to a built-in general consent form in the tutorial.
• No foreseeable risk to participants.

## 3. The Space of Experiments

The Morality Game offers large generality within its domain, making it a near-universal tool for hosting experiments in behavioral game theory and beyond. It can facilitate any experiment that involves social dilemmas expressible as game trees. This versatility stems from the inherent flexibility of game trees to represent a wide array of scenarios, ranging from bacterial competition and predator-prey relationships to marital dynamics, international rivalries, and even hypothetical first contact with extraterrestrial beings (Kurzgesagt, 2021; Liu, 2008). Morality Game experiments exploring social preference learning and higher-order Theory of Mind demonstrate this versatility.

However, other platforms exist that offer much broader capabilities than the Morality Game. For example, oTree is an open-source software tool designed for experimental economics. It supports a wide variety of field, lab, and online experiments, including multiplayer interactions such as auctions and prisoner's dilemmas. Once downloaded as a Python library, oTree provides users with many built-in features that help them program a wide range of experiments, including the ability to run studies on a web browser across various devices, test studies with artificial agents, and get real-time experiment feedback via a progress monitor. Although typically within the domain of experimental economics, oTree permits a much larger space of possible experiments because it does not standardize the user-interface, experimental procedure, or data format. For instance, otree.org provides highly diverse demo experiments ranging from the Iowa Gambling Task and a mock Twitter study to an experiment where participants annotate an image, a continuous-time public goods game, and a multiplayer word search (Chen, Schonger, & Wickens, 2016).

Similarly, PsychoPy caters to almost any single-player stimulus-response experiment in psychology. It includes a user-friendly interface called Builder, which allows researchers to design experiments by dragging and dropping stimuli, feedback, and response collection periods onto a timeline. Originating from psychophysics, PsychoPy is optimized for cognitive psychology tasks such as digit span, flanker, Stroop, go-no-go, and $n$-back. Pavlovia.org showcases PsychoPy's flexibility with demos like the balloon analogue risk task, which involves pumping up a red balloon as full as possible before it pops, whereupon the participant loses their payout. The color of the balloon, the rate by which it expands, and the on-screen background are entirely customizable, but this freedom to vary experiments along an incalculable number of dimensions makes the results of these experiments incomparable. oTree and PsychoPy catalyze research in experimental economics and psychology, by dramatically reducing the overhead time and energy devoted to programming experiments from scratch, but the infinity of studies they permit is unknowable, unmanageable, and unconstrained. This freedom is an asset and a challenge, and it places these platforms within a niche apart from the Morality Game.

The Morality Game permits a smaller infinity of experiments—a precise, controlled, manageable infinity where every study acts as a condition within an ongoing mega-experiment. The differences between Morality Game studies, and conditions within them, are based on dimensions that are both mathematically precise and psychologically meaningful. This means that even a random configuration of dashboard settings will likely produce a viable experiment with valuable data. Nonetheless, the space of possible Morality Game experiments is too vast to practically explore via brute force, meaning that discerning researchers must pick the conditions that are especially intriguing.

All Morality Game studies are conditions within a well-controlled mega-experiment because all Morality Game studies consist of a series of game trees[4]. Any series of game trees can be concatenated into a single large tree, and all trees are points in a Euclidean space of independent mathematical dimensions, which correspond to morally and psychologically relevant factors of social dilemmas. This means that any two trees have distinct sets of coordinates, allowing for the computation of the distance between them using various metrics, such as Euclidean distance. For instance, consider the simplest game: a 2-player binary dictator game, where player 1 (the chooser/dictator) chooses between option A, which has a payoff for them and player 2, and option B, with different payoffs. Players cannot communicate, share, or reverse choices, and the rules are common knowledge. A possible payoff structure might be A: $(3, 3)$, B: $(5, 1)$, where the numbers represent rewards for players 1 and 2, respectively. This game can be measured along dimensions such as the payoff difference for the self $\left(\pi_i^A - \pi_i^B\right)$, the payoff difference for the other $\left(\pi_j^A - \pi_j^B\right)$, and the payoff difference between self and other in option A $\left(\pi_i^A - \pi_j^A\right)$. These dimensions have simple mathematical definitions and map onto psychologically salient features: $\pi_i^A - \pi_i^B = 3 - 5 = -2$ (self-interest), $\pi_j^A - \pi_j^B = 3 - 1 = 2$ (altruism), and $\pi_i^A - \pi_j^A = 3 - 3 = 0$ (social comparison). More complex trees can be plotted along the same dimensions, as these equations can adapt to $n$-player $n$-option $n$-level game trees. $\pi_i^A - \pi_i^B$ and $\pi_j^A - \pi_j^B$ are the expected payoffs for the self and the other after choosing option A minus the expected payoffs before choosing option A[5], $E(payoff_i)_t - E(payoff_i)_{t-1}$. $\pi_i^A - \pi_j^A$ is the difference between the payoff for the self and the mean payoff for everyone in option A.

The Morality Game leverages mathematically precise and psychologically meaningful dimensions to describe all possible game trees and experiments. Key dimensions like payoff difference for self $\left(\pi_i^A - \pi_i^B\right)$, payoff difference for other $\left(\pi_j^A - \pi_j^B\right)$, and self-other disparity $\left(\pi_i^A - \pi_j^A\right)$ capture crucial aspects of decision-making (Stanley, Zhang, & Lewis, Inferring hidden motives: A Utility Bayesian Model of learning the values of others, 2025). Beyond payoffs[6], the Morality Game includes dimensions describing the branching factor (expected number of children per node), variance in branching factors (structural asymmetry), and total number of nodes. These factors define the optionality of the situation (aka. complexity or entropy), which significantly influence cooperation rates. Additionally, dimensions like agency, the amount of control a player has over outcomes, and truthfulness, the discrepancy between a player's beliefs and reality, are crucial for moral responsibility attribution[7] (Rudy-Hiller, 2025).

---

[4] If the series of trees in a study varies for each participant, as they typically do, then each experiment forms a constellation of trees within this space, rather than a single point.

[5] Assuming ignorance of the chooser's future choice and motivations, as if the chooser is a chance node.

[6] $\pi_i^A - \pi_i^B$ and $\pi_j^A - \pi_j^B$ are combined into a dimension called *alignment*. $\pi_i^A - \pi_j^A$ is also called *disparity*. Another payoff dimension is the total value of the payoffs or how large they are, called *stakes*. Additionally, researchers can specify how many players are involved, called $n$ *players*, which can influence diffusion of responsibility among other things (Darley & Latané, 1968). Moreover, the sequential nature of game trees allows them to encode causation and test for attributions of moral responsibility (Stanley, Responsibility Shielding: When Causally Proximate Agents Deflect Blame for Distal Enablers, 2026b).

Researchers can either build a tree and let the system find its coordinates or specify the coordinates and let the system build the tree. The algorithm constructs trees from any coordinates in constant time due to the independence of dimensions and the absence of voids in the space of trees[8]. Some dimensions are bounded (e.g., agency and truthfulness fall within [-1, 1]), and others are discrete (e.g., number of players must be an integer). Within these constraints, all locations in the space are viable[9]. For simple two-player binary dictator games, there is a one-to-one mapping between trees and coordinates. For more complex trees, multiple trees may correspond to the same coordinates. The algorithm defaults and randomizes to produce one tree per coordinate set, as the psychological differences between trees with the same coordinates are assumed to be negligible. This approach enables the system to generate trees rapidly based on the researcher's specified probability distributions, facilitating near-instantaneous tree creation before each round.

Because game trees and Morality Game experiments composed of trees vary along independent, well-defined psychological dimensions, it is possible to create studies with adjacent coordinates, differing by only one factor. This property nearly guarantees that any experiment created at random will generate data worth saving, which is why all data from formal studies is permanently consolidated within the database. While it would be an exaggeration to say that a researcher could expect to publish based on data from a randomly configured experiment, this data can reasonably be expected to enhance scientific understanding of cooperation by increasing the resolution of the cooperation landscape. In contrast, oTree and PsychoPy provide researchers with the tools to create almost any computerized experiment, akin to giving them building blocks or Legos capable of constructing any basic shape or structure. However, just as a random arrangement of Legos usually makes a mess, a random series of inputs into oTree or PsychoPy can result in a useless, unworkable experiment. By the same analogy, the Morality Game offers a kit for producing custom model airplanes: the kit includes wings, propellers, fuselages, and other components necessary for airplanes, but airplanes only. Regardless of the components chosen, the model airplane will always be viable.

Despite the viability of most Morality Game studies, the space of possible experiments cannot be exhaustively explored via brute force, necessitating researchers to be creative and methodical. This is not merely because the space of possible experiments is combinatorially explosive, but also because combinations of trees and tree parts yield emergent properties at both individual and group levels. To explore this space effectively, researchers must understand the underlying concepts and be deliberate in their experimental design. For example, binary dictator games involve simple decisions between win-win versus lose-lose options when payoffs are aligned, or trade-offs between self-interest and the interests of others when payoffs are misaligned. Adding one more choice to this simple dilemma can create a scenario participants recognize as involving trust (Alós-Ferrer & Farolfi, 2019). In a binary trust game, player 1 chooses to trust player 2 or "play it safe." Trusting player 2 is risky because it gives player 2 the opportunity to honor or betray that trust, affecting player 1's payoff. Trust thus emerges as more than the sum of the game tree parts. Similarly, adding another choice to the game can invoke second-order Theory of Mind, and simultaneous choices often create an indefinite guessing game between players. Additional psychological phenomena, such as moral shielding, can only be detected by juxtaposing two game trees. Other examples include the evolution of cooperative norms in repeated games (Axelrod, 1981; Axelrod, 1984; Nowak, 2006a; 2006b), the impact of punishment and reward systems on cooperation in many-level trees (Ghidoni & Suetens, 2022; Bru, Cabrera, Capra, & Gomez, 2003), and the role of signaling and screening (Spence, 1973; Stiglitz, 1975).

---

[7] The mapping of these factors onto psychologically salient dimensions is generally intuitive but validating each mapping could be explored in future experiments. For example, the concept of agency is tied to moral responsibility because one cannot be held accountable for inevitable outcomes ("ought implies can"). Similarly, truthfulness affects responsibility as one cannot be blamed for unforeseeable outcomes, aligning with discussions on free-will and determinism. For instance, if Makoa unknowingly transmits a disease to Akiko, he cannot be blamed unless he knew or should have known about the disease.

[8] The algorithm's construction of trees involves two main steps: generating the tree structure (linear time complexity) and assigning attributes like payoffs (constant time complexity).

[9] An exception exists when the branching factor is 1, creating a degenerate case where every parent has only one child node, eliminating options and resulting in zero agency.

These examples illustrate the richness of emergent properties within the Morality Game framework, highlighting the necessity for careful and thoughtful experimental design.

Moreover, this space of experiments is expanding without rendering previous experiments obsolete. The Morality Game can incorporate new features and factors while keeping new experiments commensurate with old ones. For example, we plan to augment the platform with emotional expression. Expressions of social emotions like anger, sympathy, and shame play a pivotal role in social exchange, despite being abstracted from the traditional trees of game theory. By boiling down a panoply of factors into a handful of mathematical objects—players, payoffs, probabilities, states, and state transitions—traditional game trees capture the essence of social dynamics but lose resolution along the way. Many nuances can be added back one-by-one once the interactions of the existing parts are well understood. We plan to allow players to express degrees of surprise and emotional valence towards other players. For instance, one player may smile or frown at another player. Between smiling and frowning exists neutral emotion, which serves as the connection point to all previous experiments where emotional expression was neutral or nonexistent. This continuity allows data from previous experiments to be seamlessly integrated with new experiments, maintaining the integrity and comparability of the overall dataset.

# 4. Platform Architecture

This section provides a concise, non-technical overview for researchers curious about the inner workings of the Morality Game. Understanding the platform's architecture empowers researchers to use the system more effectively. By grasping how the Morality Game operates, researchers can better design experiments and interpret results with confidence. For those interested in more detailed technical specifications, comprehensive documentation is available on moralitygame.com. This section covers the overall architecture, game trees, belief representation, artificial agents, and the matching function.

### 4.1 Overall Architecture

The Morality Game relies of TypeScript with React on the frontend, Python with Flask on the backend, Socket.IO for routing frontend-backed communications with web sockets, runs on Amazon Web Services EC2 servers, and stores data in SQL databases. The frontend is the part of the program that runs on individual users' local devices. If users' devices are the spokes, the server and backend is the hub, where the most intense computation takes place and where the data is saved. All user-generated signals, like responses, are routed through the server before being sent to other users, preventing direct peer-to-peer transmission. Web sockets are persistent bidirectional communication ports that relay user-generated signals to the server or server-generated signals to the users. This bidirectionality is critical because many rapid signals originate from the server, being scheduled by a clock, such as notifying players that they have abdicated responses, and the game has progressed to the next node. Mirroring the nested structure of game trees, the database stores experiment results hierarchically[10].

The remaining data is stored in a relational SQL database, which is structured like tables, with columns and rows. This remaining data includes information about players, such as anonymized player UUID, username, password, avatar shape, avatar color, PayPal ID, current study, and tutorial number; and information about experiments, such as experiment code, experiment type (tutorial, practice, pilot, simulation, official), status (configured, open, ongoing, completed, aborted, etc.), experimenter UUID, cents per payoff, number of humans, number of bots, tutorial prerequisites, and other settings. Players are identified by an anonymous UUID, and experiments are identified by a random code. There is a many-to-many mapping between players and experiments, meaning that one player can participate in may experiments and one experiment can include many players. Thus, there is a players table, an experiments table, and a players-experiments table, which stores data unique to every player-experiment pair, such as the completion URL or earnings and participation status (enrolled, active, completed, quit, etc.).

The backend is structured hierarchically, with a singleton object, called the experiment manager, that orchestrates experiments and responds to other signals from the frontend. There are multiple experiments handled by the experiment manager, multiple rounds within each experiment, multiple game rooms within each round, and usually multiple players within each game room.

Once started, an experiment cycles through each round via a while loop. The researcher specifies the slope and midpoint of a sigmoid probability distribution over the number of rounds, where the probability of continuing to the next round is initially $1$ but eventually begins to drop. The system generates a random number. If the number is greater than the continuation probability for that round, then the experiment shuts down. Otherwise, it continues. Once a round begins, the experiment emits a general update to all players and the matching function establishes which players will be grouped in that round, thus generating an array of game rooms. Next, the system sends one game tree to each game room, either selecting these trees from a list provided by the researcher or by researcher-set probability distributions over various cooperation factors: agency, truthfulness, alignment, self-other disparity, stakes, branching factor, asymmetry factor, $n$ nodes, $n$ players, and the maximum and minimum payoff value. Then the array of game rooms—one game tree per room is fed to the round manager.

[10] Planned Feature: This will allow researchers to query data across all experiments based on cooperation factors.

Like the experiment cycling through rounds, the round manager cycles through each game room while the players have yet to finish and there is still time remaining for that round. As the round manager passes by each game room, it informs the game tree in that room of the current time. Given the time, the tree either allows players more time to respond or forces the game to progress to the next node if anyone abdicated—failed to respond in time. Once the experiment finishes, it saves its status in the database, saves the participation completion statuses in the database, emits the completion URLs or PayPal payment information, and self-terminates.

**4.2 Game Trees**

Game trees are the fundamental units of the Morality Game. They are dynamic, auto-updating, self-correcting objects that generate the in-game user interface and serve as the strategic environments for each game. Trees store all in-game data and responses, which is why the raw data from each experiment is just a list of game trees, plus the experiment settings.

Trees are nested dictionaries, meaning hierarchical mappings between key-value pairs—dictionaries within dictionaries within dictionaries. Each node stores an identifier; a link to its parent; a link to the root node; the start and end times for participant responses; the payoffs; a probability, used if its parent is a chance node or to determine the next node if a choice is abdicated; the on-screen $x$ and $y$ coordinates; the identity of the chooser(s); the identity of the predictor(s); an array of choices, if submitted; an array of predictions, if submitted; the information set the node belongs to; a special object which stores player beliefs; and an array of child nodes, if any. The root node stores the identifier of the current node; the tree title; the room and round numbers; a link to the experiment object; absolute timestamps for when the round started, the round will end, the tree was emitted, etc.; an array of player objects, which stores information about the players; an adjacency matrix of pairwise matching probabilities, used by the matching function; a dictionary that directly indexes nodes; and more.

Trees are dynamic. When new nodes are added or removed from existing trees, the tree updates various attributes accordingly. For example, the $x$ coordinates of sibling nodes automatically shift to accommodate the new structure. Additionally, the tree ensures that its list attributes on all nodes are the length of the number of players. For instance, if a third player is added to a two-player tree, the payoff attributes of every node extend by one, adding a default payoff of zero for the new player. This dynamism is also evident in the trees' error-correcting mechanisms, particularly in preventing misapplied responses.

Choice and prediction responses are structured as dictionaries containing the main data—selected node identifier, key pressed (or mouse or touch event), reaction time down, reaction time up, and a timestamp—along with metadata that helps the system direct the response to the correct node on the correct tree. This metadata includes player UUID, player number, round number, room number, and node ID. Once sent from the frontend, the response is routed to the experiment manager, which forwards it to the appropriate tree if found and if the response is correctly formatted. The tree processes the response internally through a "response to tree" method. This method handles various contingencies, such as whether the response was sent on time, sent to the current node, sent to a chance node, is a choice or prediction, is a duplicate, or was sent from the wrong player. If any error is detected, the tree emits an error message to the frontend detailing the reason for the failure. If no error is detected, the tree applies the response and converts the response arrays of all nodes at the same level to immutable datatypes, preventing misapplication of responses. Once all players have responded or the time has expired, the tree updates the root node's current node ID attribute and emits a JSON file of itself to the frontend so the game can progress. This sophisticated error-handling mechanism ensures the integrity and reliability of the data collected.

**4.3 Belief Representation**

Interestingly, the Morality Game represents beliefs, including false beliefs, bridging the fields of behavioral game theory and Theory of Mind[11]. Behavioral game theory, a subfield of experimental economics, places participants in game-theoretic dilemmas, using some analytical tools from mathematical game theory while discarding unrealistic assumptions of perfect rational self-interest. Theory of Mind, a subfield of psychology, studies the ability to understand the minds of others, including their beliefs, desires, intentions, and emotions. Though these fields have limited interaction, they complement each other well. The trees of game theory provide a rigorous framework for testing and modeling beliefs about desires (payoffs), intentions (future choice probabilities), and beliefs (required for predicting outcomes of sequential choices) (Stanley, Strategic reasoning under pressure: Testing heuristics in higher-order theory of mind, 2026a). The temporal structure of game trees is conducive to modeling beliefs about the past, present, and future, whether actual, possible, or counterfactual.

In addition to storing the objectively true strategic environment, Morality Game trees represent beliefs about players, payoffs, and probabilities—whether true or false, for self and others—and can nest these beliefs to create higher-order beliefs. For example, in a binary dictator game where player 1 believes the payoffs are $\mathrm{A}\text{: }(3, 5), \mathrm{B}\text{: }(5, 1)$ when they are actually $\mathrm{A}\text{: }(5, 1), \mathrm{B}\text{: }(3, 5)$, and player 2 knows the true payoffs and knows that player 1's beliefs are incorrect, this belief is stored in a special belief dictionary. This dictionary might look like $\{\text{“@n0}{\sim}\text{p2b:@n0}{\sim}\text{p1b:@n1}{\sim}\text{payoffs”: }(3, 5)\}$, meaning that, “At node 0 player 2 believes that at node 0 player 1 believes that at node 1 the payoffs are $(3,\ 5)$.” Similarly, $\{\text{“@n0}{\sim}\text{p1b:@n9}{\sim}\text{p2b:@n1}{\sim}\text{probability”: }0.9\}$ means that “At node 0 player 1 believes that at node 9 player *will* believe that at node 1 the probability *was* $0.9$.” The key point is that this belief representation system is general enough to store beliefs for any player at any node at any time. Artificial agents base their choices and predictions on their beliefs, not the objective truth of the situation. Similarly, the user interface displays the player’s beliefs about the situation, not the objective truth stored in the code. The user interface depicts beliefs within clickable thought bubbles, allowing users to explore these beliefs without overcrowding the screen.

[11] Planned Feature: Nested belief representations were built into the Morality Game backend from the beginning. Displaying these representations on the user-interface is a planned feature.

### 4.4 Artificial Agents

Robotic agents respond based on a strategy profile, which the researcher can specify directly or indirectly by setting parameters that establish their 'personalities'. A concept borrowed from game theory, a strategy profile establishes the probability of every prediction and choice at all nodes of the tree (Von Neumann & Morgenstern, 1947). These probabilities can be directly set to 0 or 1 for deterministic responses or somewhere in between for probabilistic responses. When the researcher opts to give the bot a personality, they specify the following parameters: value for self $V_{ii}$, value for others $V_{ij}$, normal envy $\alpha_{ij}$, reverse envy (guilt) $\beta_{ij}$, risk preference $\gamma$, loss preference $\lambda_{ii}$, reasoning depth $k$, temporal discounting $\delta$, and reaction times, which are faked to make the bot appear human. These parameters are fed into a utility function, which computes the utility (happiness points) the agent derives from all possible states in the tree. Bots always choose the highest utility option if the opportunity cost is significant, but when multiple options have similar utilities, choices become probabilistic. The utilities of options are fed into a choice probability function, which returns probabilities for each available option. These choice probabilities are computed in advance of each round and stored in the strategy profile.

Researchers specify the objective parameters that determine the behavior of individual agents. Like the objective truth that players never directly see, bots never know their objective parameters. However, they will use[12] Bayesian updating to infer the parameter values of themselves and others (Stanley, Zhang, & Lewis, Inferring hidden motives: A Utility Bayesian Model of learning the values of others, 2025). They store their beliefs within a special dictionary, which acts as a centralized memory bank for all bots. Like the beliefs about tree attributes, {"1b:2b:3b:4loss": 1.2} means that player 1 believes that player 2 believes that player 3 believes that player 4 has a loss aversion parameter of 1.2. Additionally, this special dictionary stores how much each player values all other players and themselves, and beliefs about these values. For instance, {"1v2": 1.1} means that player 1 values player 2 at 1.1, {"2v1": 0.1} means that player 2 does not reciprocate this feeling, {"3v4": -0.5} means that player 3 dislikes player 4, and {"4b:3v4": -0.5} means that player 4 knows this.

To mitigate a combinatorial explosion of parameter values for every player pair and every player's belief about every other player's beliefs, this special dictionary returns any parameter that has been explicitly stored. If it has not been stored, it defaults to returning the value at the next lowest level until it reaches the objectively correct value. For example, if "1b:2b:1b:2b:1b:2v1" was not explicitly stored, it will return the value for "1b:2b:1b:2v1", and if that was not stored, it will return the value for "1b:2v1".

Every time a parameter is stored, it is timestamped, allowing researchers to chart the time course of these values, such as the building of trust between two agents. This capability enables researchers to chart the evolution of the network topology of affiliations between agents.

---

[12] Planned Feature: Bots will automatically infer the parameter values of human participants in real time. They will use these to compute utilities and choice probabilities at future game states. Ultimately, the goal is to infer the utility that each participant derives from all real and possible game states so that this data may be used to maximize utility in the real world.

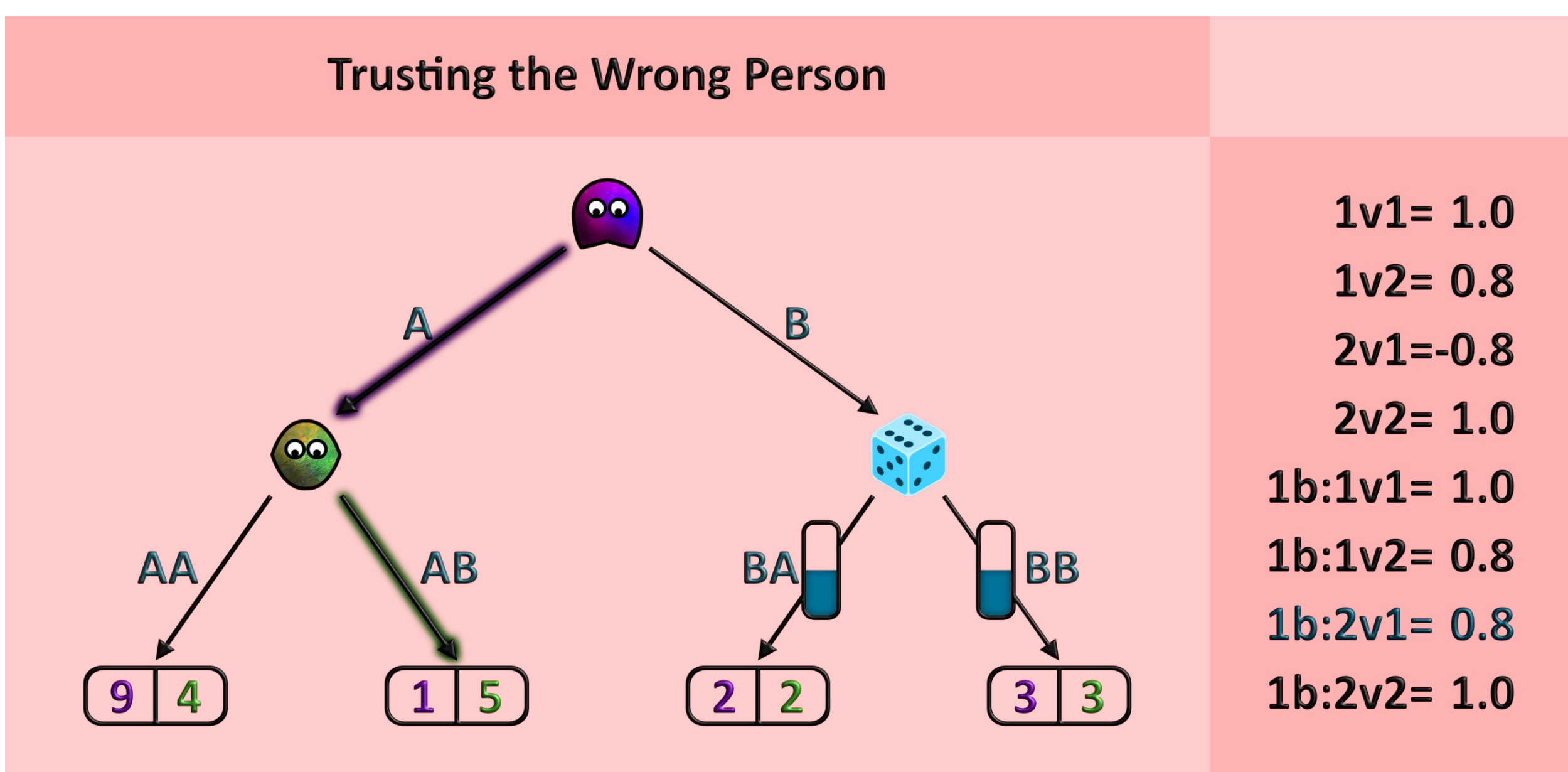

Figure 4: False Beliefs in Game Trees:

As described in the main text, this figure displays a game tree similarly to how it is represented on the Morality Game user-interface, minus the option labels (i.e. 'A', 'B', 'AA', etc.). The left panel represents the beliefs of the artificial agents, which are not made visible on the UI, but which determine how the bots behave. In this case, the purple bot trusts the green bot by selection option A on the left, expecting that the green bot will reciprocate this trust by choosing the welfare-maximizing AA on the left. Instead, the green bot betrays this trust by selecting option AB on the right, which compared to option AA, increases the green bot's payoff by 1 while decreasing the purple bot's payoff by 8. The blue text on the left panel highlights the purple bot's unwarranted trust. 1b:2v1 means that player 1 (purple) believes that player 2 (green) values player 1 at a level of 0.8. In reality, player 2 values player 1 at a level of negative 0.8, indicating dislike, which motivates the betrayal. This system of recursive belief representation can be extended to model beliefs about preferences, payoffs, probabilities, players, and how these attributes may differ from nodes in the past, present, or future, factually or counterfactually.

In the game tree above, player 1 (top, purple) trusts player 2 (left, yellow-green), who then betrays player 1 due to player 2's negative valuation of player 1. However, player 1 falsely believes that player 2 values them positively. Figure 4 illustrates how these parameters and beliefs are represented and how they influence the behavior of bots.

In the text on the right, 'v' denotes 'values,' 'b:' denotes 'believes that' '1' represents player 1, and '2' represents player 2. Both players value themselves positively at a level of 1, and both are aware of this fact about themselves and each other. Player 1 mistakenly believes that player 2 values them at a level of 0.8, shown in blue, whereas player 2 values player 1 negatively at -0.8.

From player 1's perspective, it seems rational to choose the left option, expecting that player 2 will reciprocate the trust by choosing the left option, resulting in player 1 receiving a payoff of 9 and player 2 receiving a payoff of 4. This outcome appears better for both players compared to the expected payoffs following the right-side choice. However, due to player 2's hatred, the utility calculation for player 2 favors the right-side choice, leading to betrayal:

$$2v1(AA_1 - AB_1) + 2v2(AA_2 - AB_2) < 2v1(AB_1 - AA_1) + 2v2(AB_2 - AA_2)$$

$$-0.8(9-1) + 1(4-5) < -0.8(1-9) + 1(5-4) = -7.4 < 7.4$$

$$p(choose\ AA) = \frac{e^{U(AA)}}{e^{U(AA)} + e^{U(AB)}} = \frac{e^{-7.4}}{e^{-7.4} + e^{7.4}} \approx 3.726 \times 10^{-7}$$

Robotic agents choices are set by the SoftMax choice probability function, shown above, which ensures that probabilities fall between 0 and 1, even with negative utilities, and gradually increases the chance of choosing the higher utility option as the opportunity cost grows. The opportunity cost is the utility difference between a player's best and next best option. In this case, player 2's opportunity cost is the utility of option AB (7.4) minus the utility of option AA (-7.4). This disparity is large enough that the probably that player 2 will choose option AB is very close to 1.

**4.5 Matching Function**

Rather than matching participants randomly each round, researchers can specify a probability distribution over matching probabilities. The system draws from this distribution to set pairwise matching probabilities between players, which is displayed on-screen as a possible cooperation factor. For example, some people do not tip servers at restaurants they never anticipate returning to, while others might pick up hitchhikers even though they never expect to meet again. The matching function allows researchers to investigate how matching probabilities influence participants' behaviors.

Like all distributional settings on the researcher dashboard, the user selects a distribution type—uniform, linear, normal, or sigmoid—and determines the parameters. For instance, a researcher can select a linear distribution that creates an experiment where matching probabilities tend to be large, meaning once-matched players are likely to stay matched, or where matching probabilities tend to be small, creating many one-shot encounters. Once matching is determined before a round begins, the system sets the pairwise matching probabilities between these matched players, indicating the likelihood they will meet each other in the subsequent round.

Matching probabilities are stored in an adjacency matrix, where all players in a game room have a row and column. For example, in a four-player game, the cell in the third row and fourth column stores the chance that player 3 will meet player 4 in the next round. The system samples from the researcher-specified matching probability distribution and saves the values in the adjacency matrix. Since the probabilities in each game tree/room are interdependent (e.g., if $p(\text{match } 1\&2) = 1$ and $p(\text{match } 2\&3) = 1$, then $p(\text{match } 3\&1)$ must also be 1), the system adjusts for these interdependencies to display accurate matching probabilities on each user's interface.

Although this system determines and displays matching probabilities one round at a time, it does not account for pairwise matching probabilities across an entire experiment, which also depend on the expected number of rounds and the number of players in the study. However, researchers can opt for a matching system that establishes constant matching probabilities across the experiment. Instead of using adjacency matrices for individual game rooms, this system generates an adjacency matrix for all players in the experiment across all rounds. The downside of this approach is that it cannot guarantee the approximation of these probabilities if a player drops out, and it can only establish pairwise matching probabilities, not probabilities over trios or larger groups, making it suitable only for two-player games.

**Platform Comparisons**

**oTree and PsychoPy** (Chen, Schonger, & Wickens, 2016)**:**

oTree and PsychoPy are highly versatile platforms capable of creating a wide variety of experiments. oTree excels in hosting multiplayer experiments and requires coding in Python, while PsychoPy allows experiment design via both code and a GUI. In contrast, the Morality Game does not require any coding and provides a well-controlled experimental environment where even randomly configured settings can yield valuable data.

**Breadboard** (McKnight & Christakis, 2016)**:**

Breadboard shines in creating and analyzing social networks with diverse configurations in online multiplayer games using MTurk participants (Amazon Mechanical Turk). It uses a scripting dialog and a user interface to show network graphs, making it a hybrid between coding and a GUI. Breadboard allows more complex network interactions than the Morality Game, which currently partitions participants into discrete game rooms. However, the Morality Game offers more controlled stimuli, enhancing experimental rigor. Future updates aim to integrate network analytics and more sophisticated matching functions to bridge this gap.

**Gambit and Game Theory Explorer (GTE)** (Savani & Turocy, 2025; Savani & Stengel, 2015)**:**

GTE focuses on traditional mathematical game theory, providing tools for exploring game theory games via a GUI without coding. They cater to mathematicians and students, not human subjects research, and assume perfect rationality and self-interest in players. In contrast, the Morality Game is designed for behavioral experiments with real people, aligning more with behavioral game theory. Both platforms can serve educational purposes.

**Open Science Framework (OSF)** (Center for Open Science, 2025)**:**

OSF is an open-access repository dedicated to transparency, reproducibility, and collaboration. The Morality Game shares these goals and plans to link to OSF via its API, allowing researchers to store research materials on OSF while accessing them within the Morality Game. This integration enhances data management and collaborative efforts, providing support throughout the research lifecycle.

**Arcade Learning Environment (ALE)** (Bellemare, Naddaf, Veness, & Bowling, 2013)**:**

The ALE assimilates old Atari video games for reinforcement learning research, providing common benchmarks for AI training. Similarly, the Morality Game consolidates research around standard metrics, producing structured data suited for AI training. Planned features include allowing researchers to upload custom models (bots) to fit human data. These models will be scored on their predictive accuracy, creating an ongoing competition that dynamically updates with new data, akin to an ongoing competition between models.

## 5. Future Directions

The Morality Game is a catalyst for researching how to increase cooperation. It catalyzes research by directing researchers to converge upon a single ongoing mega-experiment. To be more effective, the Morality Game must expand the scope of that experiment, foster greater connections between researchers, and evolve into an integrated ecosystem that supports the entire research lifecycle, from participant recruitment to data analysis and dissemination.

The space of Morality Game experiments, while vast and mathematically precise, must grow to encompass the richness of the offline human social world, which is filled with complexities such as laughter, facial expressions, social roles, and nuanced interactions. The difference between the controlled environment of the Morality Game and the real world is quantifiable and can be bridged by incrementally incorporating these real-world elements into the platform.

For instance, facial expressions are central to human cooperation, yet the avatars on the Morality Game do not currently blush, smile, or sneer. Adding a webcam feature to enable live video calls would simultaneously introduce variables such as gender, race, attractiveness, emotion, and natural language—an incalculable number of uncontrolled factors. Instead, the Morality Game will incrementally introduce variables that are mathematically precise and aligned with psychologically validated constructs.

Initially, we plan to allow bots and users to express degrees of emotional valence and surprise. This will be done through controlled, quantifiable measures. Players will be able to communicate sentiments about specific game tree attributes by attaching emotional indicators—such as happy/sad or surprised/unsurprised—to relevant icons. For example, positive valence toward a player might convey, "I like you," while negative valence toward a past node could indicate, "I was unhappy with that outcome." Combining mild positive valence and non-surprise towards one of a counterpart's options might signal, "I want and expect you to choose this option," while negative valence and surprise towards an alternative could mean "...and I'll be very disappointed if you choose this option instead."

This structured social language will be composed of fundamental elements: valence (good, bad, happy, sad) and belief (true, false, probable, improbable), applied to existing game tree parts—players, payoffs, probabilities, states, and state transitions; past, present, future, factual, and counterfactual. These basic components can form machine-readable sentences like "I like you," "I regret my choice," and "I'm surprised that you believed that they would be willing to make that choice," all of which can be depicted pictographically on the user interface[13].

[13] Many emotions result from combinations of these atoms: excited = positive valence to future state, fearful = negative valence to future state, sadness = negative valence to past state (especially if caused by chance node), anger = negative valence to past state knowingly caused by other agent, frustration = lack of good options, guilt = negative valence to self (regardless of being watched), shame = believed negative valence to self in the minds of others, and anxiety = high entropy/too many options (made worse by uncertainty from chance events and especially ambiguous probabilities). However, some fundamental emotions cannot be expressed in these game theoretic terms, such as disgust, awe, and humor.

Step by step, the Morality Game will develop stimuli that capture additional variance in cooperation rates, thereby enhancing its ability to inform public policy, institutional design, and personal relationships. By methodically introducing these new elements, the platform will better reflect the complexities of real-world social interactions, making its findings more generalizable and impactful[14].

The Morality Game must incorporate many other cooperation factors. Among these are different matching arrangements, prior connection strengths, and relational mobility. The matching function will be augmented to include all of these. Participants can be arranged into social networks with varying degrees of homophily, clustering coefficients, and degree assortativities. Eventually, participants will choose their own partners, perhaps even sending invitations, like a dating app. Additionally, the Morality Game will allow manipulation of player identities, enabling players to wear masks, adopt alternate forms, and change characteristics and group affiliations.

To better unite interested scientists, the Morality Game will add social networking features and tools for disseminating their findings. Leveraging the API from the Open Science Framework, researchers will be able to upload their preregistrations, pre-/post-prints, and papers to their data galleries. This will function like a public profile page on a social media site fused with an academic journal, where researchers can post platform-specific papers, datasets, and interactive data visualizations, both automatically generated by the system and custom-made by the researchers. Hyperlinks within and between Morality Game papers will connect ideas, fostering a rich web of interconnected research.

To become a nexus for research, advancing many aspects of the research lifecycle, the Morality Game will garner a large pool of participants, integrate with other platforms and technologies, facilitate data analysis, host contests between computational models, and expand the types of experiments and data it can manage. Because many of the experiments are engaging, intellectually stimulating, and paid, there is significant potential to attract a large pool of returning participants. The value of data from the same participant across multiple studies is greater than that of two different participants each doing one study, as it allows for the detection of patterns over time. For example, longitudinal data can reveal how cooperation strategies evolve with repeated interactions under varying conditions.

To streamline the research process, we will add participant management and scheduling features. Additionally, expanding the scope of experiments the Morality Game can cover will involve making the platform compatible with neuroimaging studies and integrating with other platforms like Qualtrics for researchers interested in collecting survey data before or after an economic game phase of their study. Furthermore, researchers will be able to submit custom computational agents to the platform, which act as models attempting to fit the data. Every time new data is added to the database, the system will re-rank the models based on their ability to fit as much of the data as possible. The more results a model can predict without modification, the more confidence we can have in its validity.

---

[14] Salience Landscape: By analogy, a dating profile that lists height as a number makes height more salient than meeting the same person face to face, where height is equally real but competes for attention with charisma, humor, and warmth. People weight height more heavily when the medium foregrounds it as a digit, even though the person being evaluated is identical. The Morality Game risks distorting the salience landscape the same way, by drawing attention to a chosen few of the factors that are always present.

Moreover, we plan to develop the bots to dynamically learn and adjust their behavior based on their judgments of their counterparts. These agents will characterize the personalities of other players using parameters such as self-value, other-value, enviousness, and reasoning depth. They will forecast future character judgments contingent on their counterpart's choices and use this information to inform their predictions. For example, if an agent predicts its future choices will be affected by a selfish choice from its counterpart, it will account for hypothetical feelings of betrayal and retribution in its predictions.

At the system level, the platform will use Bayesian parameter estimation to infer the parameter values of human participants, facilitating analyses of individual differences and network analyses of pairwise connection strengths. Ultimately, the goal is to estimate utilities for each player at all possible states, aligning with the utilitarian aim of maximizing overall utility. Once agents can estimate participant parameters, they should be able to predict and imitate human choices, even in novel situations. This capability will allow us to explore a far larger space of experiments with imitation bots, creating a high-resolution model of the cooperation landscape.

Looking ahead, we also envision conducting studies in remote small-scale societies. With a more mobile-friendly platform, it will be possible to travel to remote indigenous communities with mobile hotspots and phones/tablets to collect economic game data, akin to the efforts of Joseph Henrich and collaborators (Henrich, 2004). This approach will help overcome WEIRD sample bias and test the ecological validity of Morality Game experiments. Running parallel lab and field studies will further validate our findings and ensure their applicability across diverse contexts.

By expanding its capabilities, fostering deeper connections among researchers, and integrating with complementary platforms, the Morality Game aims to become a comprehensive, multifaceted research hub. This vision not only enhances the platform's utility but also drives forward our understanding of cooperation and morality, providing valuable insights that can shape legal structures, institutional architecture, and personal interactions. Ultimately, increasing cooperation is a means to increasing utility, and maximizing utility is the point of the Morality Game.

***Acknowledgments***

G.N.S. is grateful to Jun Zhang and Rick Lewis for their support and mentorship during his doctoral studies at the University of Michigan. G.N.S. thanks Rose Stanley for her dedication and support throughout this work and throughout graduate school.

***Open science and data availability statement***

Data generated through the Morality Game platform are openly available and can be downloaded at https://moralitygame.com. To keep participants naive to hypotheses and stimuli, study data are gated from participant accounts. Researchers can view and download all data after logging in with the researcher password "utility." If you have any problems with the site or would like to ask questions, you may contact Greg Stanley at gregstan@umich.edu.

***Ethics statement***

The Morality Game platform and its pilot studies were approved by the University of Michigan Institutional Review Board, Health Sciences and Behavioral Sciences (IRB-HSBS; Protocol HUM00243214), and determined exempt from ongoing review. All participants provided informed consent prior to participation.

***Declaration of competing interest***

The authors declare no conflicts of interest related to this study.

***Financial support statement***

This research received no specific grant from any funding agency in the public, commercial, or not-for-profit sectors.

***Declaration of generative AI and AI-assisted technologies in the writing process***

During the preparation of this work, the authors used OpenAI's ChatGPT, Anthropic's Claude, and Perplexity as research-assistance tools for brainstorming, literature search assistance, code drafting/ debugging, analysis support, and manuscript revision. All AI-generated suggestions, text, code, and analyses were critically evaluated, substantially revised where needed, and independently verified by the authors. The authors take full responsibility for all study design decisions, analyses, interpretations, citations, and final wording.